\documentclass[runningheads]{llncs}
\usepackage{graphicx}
\usepackage{lipsum}  
\usepackage{multirow}
\usepackage{array}
\usepackage{float}
\begin{document}
\title{Deep Learning for Automatic Strain Quantification in Arrhythmogenic Right Ventricular Cardiomyopathy}
\titlerunning{Deep Learning for Automatic Strain Quantification}
%

\author{
Laura Alvarez-Florez\inst{1,2,3}$^*$,
Jörg Sander\inst{1,2}$^*$,
Mimount Bourfiss\inst{4},
Fleur V. Y. Tjong\inst{3,7}, 
Birgitta K. Velthuis\inst{5} \&
Ivana Išgum\inst{1,2,6,7}
}
\authorrunning{L. Alvarez-Florez, J. Sander et al.}

\institute{
Department of Biomedical Engineering and Physics, Amsterdam University Medical Center, location University of Amsterdam, The Netherlands\and
Informatics Institute, University of Amsterdam, The Netherlands\and
Heart Center, Department of Clinical and Experimental Cardiology, Amsterdam University Medical Center, location University of Amsterdam, The Netherlands\and
Department of Cardiology, University Medical Center Utrecht, The Netherlands\and
Department of Radiology and Nuclear Medicine, University Medical Center Utrecht, The Netherlands\and
Department of Radiology and Nuclear Medicine, Amsterdam University Medical Center, location University of Amsterdam, The Netherlands\and 
Amsterdam Cardiovascular Sciences, Amsterdam, The Netherlands \\
$^*$ These authors contributed equally to this work. }


%
\maketitle              
\begin{abstract}
Quantification of cardiac motion with cine Cardiac Magnetic Resonance Imaging (CMRI) is an integral part of arrhythmogenic right ventricular cardiomyopathy (ARVC) diagnosis. Yet, the expert evaluation of motion abnormalities with CMRI is a challenging task. To automatically assess cardiac motion, we register CMRIs from different time points of the cardiac cycle using Implicit Neural Representations (INRs) and perform a biomechanically informed regularization inspired by the myocardial incompressibility assumption. To enhance the registration performance, our method first rectifies the inter-slice misalignment inherent to CMRI by performing a rigid registration guided by the long-axis views, and then increases the through-plane resolution using an unsupervised deep learning super-resolution approach. Finally, we propose to synergically combine information from short-axis and 4-chamber long-axis views, along with an initialization  to incorporate information from multiple cardiac time points. Thereafter, to quantify cardiac motion, we calculate global and segmental strain over a cardiac cycle and compute the peak strain. The evaluation of the method is performed on a dataset of cine CMRI scans from 47 ARVC patients and 67 controls. Our results show that inter-slice alignment and generation of super-resolved volumes combined with joint analysis of the two cardiac views, notably improves registration performance. Furthermore, the proposed initialization yields more physiologically plausible registrations. The significant differences in the peak strain, discerned between the ARVC patients and healthy controls suggest that automated motion quantification methods may assist in diagnosis and provide further understanding of disease-specific alterations of cardiac motion.

\keywords{Implicit Neural Representations \and Image Registration \and Strain \and Cardiac Motion \and Arrhythmogenic Right Ventricular Cardiomyopathy.}

\end{abstract}

\section{Introduction}

Heart motion abnormalities serve as indicator of cardiac disease and its severity. For arrhythmogenic right ventricular cardiomyopathy (ARVC) patients, characterization of wall motion abnormalities is an integral part of diagnosis. In clinical practice, cardiac magnetic resonance imaging (CMRI) is considered the reference standard for the assessment of motion abnormalities \cite{scatteia2017strain}.  The assessment typically relies on visual inspection by radiologists. This process is challenging, and therefore subjective and lacks a quantitative description  \cite{qiao2020temporally}. Automated deep learning methods may offer accurate and reproducible quantification, and allow subsequent interpretation of cardiac motion. 

Unlike other methods quantifying cardiac motion and strain such as CMRI tagging, motion quantification with cine CMRI is derived as a post-processing and does not require additional imaging or long acquisition and processing time \cite{bucius2020comparison}.  Classic computation methods for motion quantification in cine CMRI include feature tracking and image registration \cite{qiao2020temporally}. More recent approaches used deep learning \cite{morales2021deepstrain,puyol2018strain,upendra2021motion,meng2022mulvimotion,qin2023generative}. These methods compute the displacement fields by mapping image intensities or anatomical landmarks across different time points within the cardiac cycle. These methods have shown the ability to perform registration fast, making them appealing for clinical use \cite{wang2020deepflash}. Nevertheless, their performance does not necessarily outperform classical registration methods  \cite{wolterink2022implicit}. Registration using Implicit Neural Representations (INRs), a distinctive recently proposed method that employs a neural network to represent the implicit transformation function between two images, outperforms previous CNN-based methods and offers fast analysis \cite{wolterink2022implicit}. Lopez et al. recently applied INR based networks to deformable image registration of CMR images \cite{lopez2023warppinn}. However, CMRI registration is challenged by image acquisition throughout a cardiac cycle that leads to misalignment between consecutive slices and low through-plane resolution.  In response to these limitations, recent studies incorporated preliminary alignment of images before computing anatomical displacements \cite{upendra2021motion,SANDER2023107266}.

To quantify cardiac motion addressing these challenges, we propose a method that performs CMRI registration between different time points in the cardiac cycle and then computes the strain.  First, we employ inter-slice alignment and unsupervised deep learning super-resolution to enhance the performance of implicit neural representations. Second, we include different cardiac views during registration. Third, our method utilizes transfer learning across different time points to incorporate temporal information into the initialization of INRs. Additionally, in line with earlier works \cite{qin2023generative}, our registration method constrains the network to adhere to biomechanically informed rules by incorporating a  regularization technique. Lastly, we compute the strain in 3D from the displacements leveraged from the proposed registration method. We evaluate the method with a set of ARVC patients and control subjects. The novelty of our work lies on the addition of super resolution along with preliminary slice alignment and the  incorporation of multiple views into INRs. Moreover, we propose a unique way to transfer the registration transformation learned by INRs from one point in the cardiac cycle to the next.

\begin{figure}[h]
\centering
    \includegraphics[width=\textwidth]{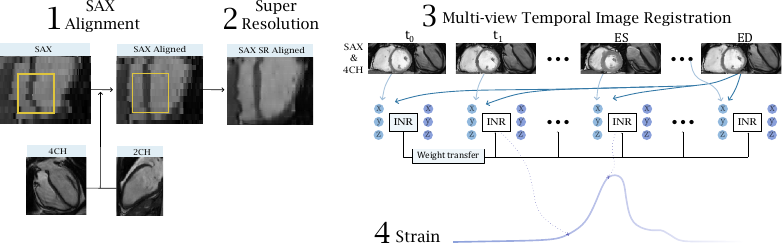} 
    \caption{ Overview of the proposed method. First alignment and then super-resolution. The SAX and 4CH views are fed into the INR. The weights from each network are transferred to initialize the succeeding one. Lastly, the strain is computed.}
    \label{fig:method}
\end{figure}

\section{Method}

 During cine CMRI acquisitions, each short axis (SAX) slice is typically captured during breath-hold. Variability in a patient's breath-hold relative to the imaging volume can result in misalignment of the SAX slices (Figure \ref{fig:method}). To address this and enhance registration performance, we first rigidly align slices in the through-plane direction of the SAX view. To guide the alignment, the 4-chamber (4CH) and 2-chamber (2CH) views are used as reference. In addition, we use segmentations of the left ventricle (LV) blood pool from the SAX, 2CH and 4CH views derived by a deep learning segmentation method \cite{sander2020automatic}. Initially, the SAX, 2CH and 4CH views are transformed to the same coordinate space.  To determine the rigid registration parameters, we optimize a translation matrix with learnable weights through an adaptive moment estimation (Adam) optimizer. To calculate the loss, first the SAX image is resampled using bilinear interpolation based on the alignment parameters. Then, we warp the aligned SAX image into each of the long-axis views (i.e. 4CH and 2CH). The computation of the loss that guides the learning process is composed of two components: the image loss and the segmentation loss. The image loss is calculated over the masked area of the myocardium. For this, the Normalized Cross Correlation loss (NCC) is combined with a differentiable version of Normalized Mutual Information (NMI) \cite{de2020mutual} loss. The NMI is then scaled by a factor to ensure both losses are of the same magnitude. The Dice loss is used to compute the segmentation loss. Both components of the loss (i.e.  image and segmentation) are calculated for the 2CH view and for the 4CH view and combined into one component.  The method is optimized for a total of 2,000 iterations with a learning rate of 0.01.

Given the highly anisotropic resolution of CMRI, increasing the through-plane resolution prior to registration of the cardiac volumes from the cine acquisition might lead to smoother displacement fields. Therefore, we employ an unsupervised deep learning super-resolution method \cite{SanderAutoencoder} that generates intermediate slices as a linear combination of preceding and succeeding slices interpolated in the latent space.  We up-sampled the volumes using a factor of 6, increasing the average number of slices from 15 to 85.

To perform registration of the different volumes from the cardiac cycle, we use INRs. INRs are trained using a subset of randomly selected coordinates from the fixed and moved images. The network employs a loss function that minimizes the discrepancy between the intensity of the warped moving and the fixed images. We use the architecture proposed by \cite{wolterink2022implicit}, the implementation consists of 3 layers, each with 256 hidden units. The network employs a periodic activation function and the Adam optimizer with a learning rate of $10^{-4}.$

To improve the performance of INRs, we incorporate information from both the 4CH and SAX image coordinates into the implicit image registration. To combine the 4CH and SAX views during registration, the coordinates are transformed into a unified canonical space. In every iteration, we forward a batch of 10,000 coordinates from the SAX and another batch of 10,000 from the 4CH view. The loss is the combined NCC from both views, enabling the network to optimize both transformations simultaneously.

We introduce an adaptation to the Jacobian regularization of INRs \cite{wolterink2022implicit}. Rather than applying a uniform regularization across the entire image, we incorporate a weighting parameter that assigns greater importance to the myocardium (MYO). The Jacobian regularization encourages the neural network to learn transformations that preserve local volumes, promoting smoother displacements that are more physiologically plausible and preventing folding. Our weighting strategy ensures that the regularization is focused on the foreground (i.e. MYO), while applying a more relaxed penalty to the background of the image. By enforcing this greater penalty on the foreground, this regularization endorses the myocardial incompressibility assumption. The overall loss $\mathcal{L}$ for our method is presented in Equation \ref{eq:total-loss}.

\begin{equation}
\mathcal{L} = \mathcal{NCC}^{sax} + \mathcal{NCC}^{4ch} + \alpha_{fg} \cdot J_{fg}^{sax} + \alpha_{bg} \cdot J_{bg}^{sax} + \alpha_{fg} \cdot J_{fg}^{4ch} + \alpha_{bg} \cdot J_{bg}^{4ch}
\label{eq:total-loss}
\end{equation}

Where $\mathcal{NCC}^{sax}$ and $\mathcal{NCC}^{4ch}$ are the NCC losses for the SAX and 4CH. The $J_{fg}^{sax}$, $J_{bg}^{sax}$, $J_{fg}^{4ch}$, and $J_{bg}^{4ch}$ components represent the foreground and background regularization terms for the SAX and 4CH views, and $\alpha_{fg}$ and $\alpha_{bg}$ are the weights applied to the regularization terms for the foreground and background.

INRs require optimizing a new network for each registration. However, we can exploit the information from the registration of cardiac volumes between earlier time points to the later time points of a cine acquisition. Hence, while requiring to train $n$ separate networks for registration of $n$ time points in the cardiac cycle, we make this process more efficient and leverage the information learned from one time point to the next. Specifically, given an initial time point $t_0$ and a final time point $t_n$, representing the start and end of the cardiac cycle at end-diastole (ED), we train a network for each image volume pair starting from $\{t_i,t_n\}$ where $i$$=$0,...,$n$-1. The optimized weights from each iteration are carried forward to initialize the succeeding network $\{t_{i+1},t_n\}$. This approach ensures that the network starts by mapping small transformations. Hence, when the network is faced with the more intricate task of mapping from contraction to relaxation,  it does not start anew but leverages a pre-established optimized setting. By employing an optimized setting as a foundation for each successive time point computation, our approach also aims to  diminish the variability in initial network states and constrains the exploratory solution space.

Lastly, to analytically calculate the deformation gradient and subsequently the strain tensor, we leverage the automatic differentiation ability of deep learning frameworks. We compute the circumferential and longitudinal components of the strain. These components represent changes in wall thickness (radial strain), and circumferential length (circumferential strain). To facilitate an interpretation of the strain that is aligned with the intrinsic physiological shape of the LV, the resulting Cartesian-based strain tensor is transformed into a polar coordinate system. In contrast to the LV, which can be approximated as a circular or ellipsoid structure, the right ventricle (RV) exhibits a more complex and irregular shape. To account for this, we compute the outward-pointing normals of the RV contour, providing a unique local direction for each voxel. These normals define the radial direction, essentially pointing outwards from the center of the ventricle. The circumferential direction is determined as orthogonal to the radial direction within the contour plane. Figure \ref{fig:method} illustrates the proposed method.

\begin{figure}[H]
\centering
    \includegraphics[width=\textwidth]{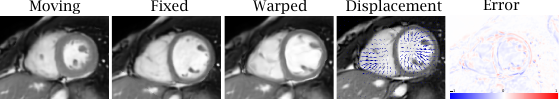} 
    \caption{Evaluation of the registration results. This results were achieved by our proposed method. The error represents the difference between the warped and fixed images. }
    \label{fig:displacement}
\end{figure}

\section{Evaluation}

\subsection{Dataset}

We evaluate our method on a dataset of conventional steady-state free precession sequences. The dataset was compiled from 47 ARVC patients and 67 control subjects. The data consists of SAX, 2CH and 4CH views captured during breath hold. Each sequence included 25 to 40 phases spanning one cardiac cycle, with a repetition time ranging from 2.6 to 3.4 ms. The end-systole (ES) and ED time points were identified by the expert radiology technicians as a part of clinical workup. Following this, segmentations for the LV blood pool, LV MYO and RV blood pool of the SAX, 2CH and 4CH slices were conducted automatically \cite{sander2020automatic}.

\subsection{Evaluation metrics}

To evaluate the proposed registration method, we compare the segmentations between the fixed and warped images using overlap and distance metrics. The overlap between segmentations is measured with the Dice Similairty Coefficient (DSC), and the distance between the segmentations' boundaries with the Hausdorff Distance (HD).  Furthermore, to evaluate the quality of the displacement fields, we calculate the determinant of the Jacobian matrix, which provides a measure of local volume changes. If the Jaccobian determinant is equal to 1, the transformation is volume-preserving. To discern the differences in strain between the ARVC patients and controls, we calculate the peak strain. This is defined as the maximum point of strain reached, representing the largest deformation experienced by the cardiac tissue.  We employ a Kruskal-Wallis test to determine whether there was a statistically significant difference between the two groups.  Significance was defined with a p-value lower than 0.05.

\section{Experiments and Results}

\subsection{Evaluation of the registration}

We performed a quantitative and qualitative evaluation of the method. We limited the quantitative evaluation to a subset of 32 patients from the dataset, and registered ES to ED. The qualitative evaluation was performed on one subject over the entire cardiac cycle (i.e. registering all the cardiac time points sequentially). The quantitative results are listed in Table \ref{tab:results-abalation}. Figure \ref{fig:displacement} shows the resulting displacement vector fields of the proposed method.

We quantitatively assessed the influence of each component in our method (i.e. slice alignment, super-resolution, multiple views, weighted Jacobian regularization and our proposed transfer learning initialization) by performing iterations with and without this steps. The results are listed in Table \ref{tab:results-abalation}.  The overall best method was achieved when including all the proposed components. The largest contribution to the DSC, for all cardiac structures (LV, MYO and RV), came from performing preliminary alignment and super-resolution along with our proposed initialization of weights. The initialization showed to have a significant positive impact in the DSC, particularly for the RV, in both the SAX and 4CH views. Including super-resolved volumes showed the largest reduction in the HD, which was further improved with the proposed initialization. When adding multiple views, there was a considerable increase in the DSC for the 4CH view. Incorporating the weighted strategy (with $\alpha_{fg}=0.05$ and $\alpha_{bg}=0.0001$) into the Jacobian regularization, compared to applying the regularization uniformly (with $\alpha=0.05$) across the image, had the largest reduction in the Jacobian determinant. 

\begin{figure}[h!]
\centering
\resizebox{0.8\textwidth}{!}{%
    \includegraphics[width=\textwidth]{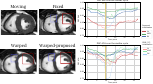} }
    \caption{
     Warped image obtained using the initialization from \cite{wolterink2022implicit}, and the one obtained with our proposed initialization (left). The boxes highlight regions from the image that are more challenging to register. Comparison of the DSC between the two initialization strategies for the same subject over all the cardiac time points (right).} 
    \label{fig:qualitative-zoom}
    \vspace{-0.23cm}
\end{figure}

For the qualitative evaluation, we registered each time point of the cine sequence to the ED, set as the fixed image. We compare our proposed initialization method, that transfers the optimization state from one time point to the next one, with the initialization from \cite{wolterink2022implicit}, which is a modified version of the Xavier initialization \cite{glorot2010understanding}. Figure \ref{fig:qualitative-zoom} shows the impact of including the transfer learning strategy to initialize the proposed method: the proposed initialization leads to a more physiologically plausible registration (left); to visually compare, we computed the DSC scores for both strategies (right). The results show that the proposed method, compared to using a standard initialization, results in a better registration performance for larger deformations (i.e. ES to ED). This is most prominent for the RV, specially in the 4CH view.


\begin{table}[t]
\resizebox{0.999\textwidth}{!}{%
\begin{tabular}{l|c|ccc|ccc|ccc|}
\multicolumn{1}{c}{}  & \multicolumn{1}{c}{HD}              
& \multicolumn{3}{c}{DICE SAX}                 & \multicolumn{3}{c}{$|$JACOBIAN$|$ -1}             & \multicolumn{3}{c}{DICE 4CH}                  \\ \cline{2-11}
Experiment                                                                                                                  & \multicolumn{1}{c}{AVG}             & LV            & MYO           & RV            & LV            & MYO           & RV            & LV            & MYO           & RV            \\ \hline

\begin{tabular}[c]{@{}l@{}}Proposed \end{tabular}
 & \textbf{10.33} & \textbf{0.92} &\textbf{ 0.79} &\textbf{ 0.87} & \textbf{0.04} & \textbf{0.04} & \textbf{0.05} & \textbf{0.92} & \textbf{0.85} & \textbf{0.90} \\

\begin{tabular}[c]{@{}l@{}}INR + WJ + AL + SR + MV \end{tabular}
 & 15.74 & 0.90 & 0.78 & 0.84 & 0.05 & \textbf{0.04} & \textbf{0.05} & 0.91 & 0.83 & 0.81 \\
 
\begin{tabular}[c]{@{}l@{}}INR + WJ + AL + SR\end{tabular}                          & 14.01 & 0.91 & 0.78 & 0.84 & 0.05 & \textbf{0.04} & \textbf{0.05} & 0.90 & 0.76 & 0.79 \\
 
INR + WJ + AL                                                                                          & 23.16 & 0.88 & 0.71 & 0.80 & \textbf{0.04} & \textbf{0.04} & \textbf{0.05} & 0.88 & 0.72 & 0.77 \\
 
INR + WJ                                                                                                      & 19.49 & 0.87 & 0.71 & 0.80 & \textbf{0.04} & \textbf{0.04} & \textbf{0.05} & 0.88 & 0.71 & 0.76 \\
 
INR                                                                                                                    & 19.46 & 0.81 & 0.70 & 0.76 & 0.17 & 0.18 & 0.22 & 0.79 & 0.71 & 0.71 \\
\hline
\end{tabular}}
\vspace{0.01cm}
\caption{Results showing the benefit of including different components from the method. INR represents the baseline registration method, weighted Jacobian (WJ) regularization, preliminary inter-slice alignment (AL), super-resolution (SR), transfer learning initialization (INIT), and multiple views (MV). The proposed method includes (INR+WJ+AL+SR+MV+INIT).}
\label{tab:results-abalation}
\vspace{-0.37cm}
\end{table}

\subsection{Evaluation of the strain}

\begin{figure}[h!]
\resizebox{0.99\textwidth}{!}{%
    \includegraphics[width=\textwidth]{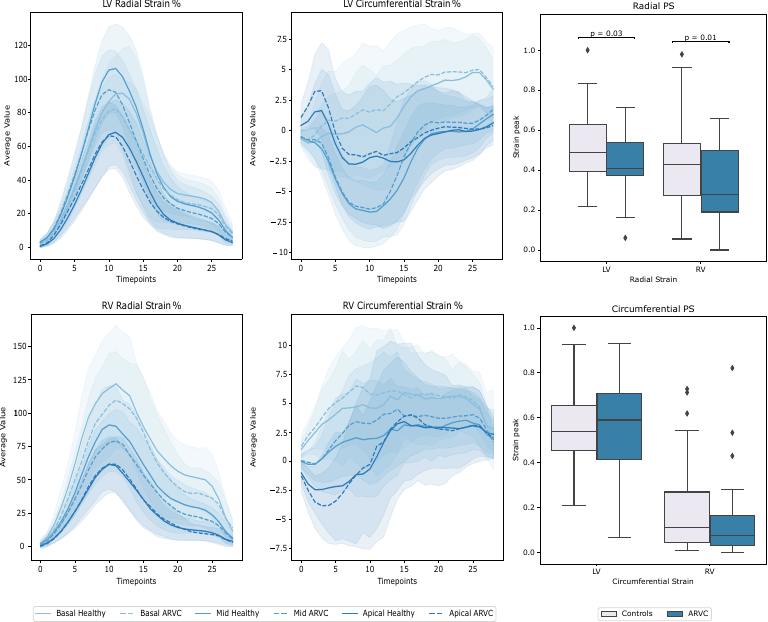} }
    \caption{Strain curves for ARVC patients and controls (left), each color represents one segment of the heart. Peak strain comparison between mid slices of the  groups (right).}
    \label{fig:strain-curves}
\end{figure}

For the evaluation of the strain, we compute the radial and circumferential strain for the basal, mid and apical slices of each volume. The derived strain curves for each segment are presented in Figure \ref{fig:strain-curves}, left. When comparing the average strain curves between the two groups, a slight reduction in the RV radial and circumferential strain is present, specially on the basal and mid slices. This decrease is not so evident for the LV strain. The normalized peak strain for each group is detailed in Figure \ref{fig:strain-curves}, right. The strain results are analogous across the board, however, a significant difference (p=0.01) in the radial peak strain of the RV was identified between the ARVC patients and the controls. Another weak significance was found for the radial LV strain.

\newpage
\section{Discussion and conclusion}

We presented a method for temporal registration of cine CMRI that includes aligned super-resolved volumes, a biomechanically informed regularization and incorporates different cardiac views and time points into implicit neural networks. The method was evaluated on a dataset including ARVC patients and controls. The results showed that the method addressed the inherent limitations of cardiac imaging through preliminary slice alignment, super-resolution and inclusion of multiple views, leading to a substantial improvement in the performance of the registration method. Our proposed initialization based on transfer learning from previous time points yielded more accurate and physiologically plausible registrations. Furthermore, adopting a weighted Jacobian regularization resulted in more realistic volume-preserving transformations. For ARVC patients, a reduction in the strain, especially in the RV, is expected. This anticipated difference in strain was confirmed upon inspection of the radial and circumferential strain values for the basal, middle, and apical sections of the heart, which revealed differences in the peak radial strain between the ARVC patients and controls. Compared to the values reported in literature, our computed baseline radial strain was on average higher. This can be due to the computation of the strain in 3D instead of in 2D, as commonly done by clinical software \cite{meng2022mulvimotion}. This discrepancy is more pronounced for our method possibly due to the computation of these values in high-resolution super-resolved volumes. Additionally, our model encountered challenges when calculating the circumferential strain for the RV, particularly for the more basal slices. This difficulty might explain the lack of observable but expected difference in the circumferential strain between ARVC patients and controls \cite{heermann2014biventricular}. A limitation of this study is the absence of comparison to established strain values obtained from clinically validated methods like tagged MRI or feature tracking. Another limitation, intrinsic to INRs, is the necessity to train a separate network for each patient and each of their respective image time pairs. However, we alleviate the process by introducing a transfer learning strategy that has the potential to speed up this process. As accuracy is enhanced by sharing weights from one time point registration to the next, fewer epochs may be sufficient to achieve accurate registration. This will be further explored and evaluated in our future research. While registration methods show promise for providing new insights in the quantification of motion abnormalities, further research evaluating clinical value of the approach is warranted.

\bibliographystyle{splncs04}
\bibliography{references}

\end{document}